\documentstyle[aps,amsfonts,epsf,pre]{revtex}\begin{document} 
\draft
\author{Magnus Rattray}
\address{Computer Science Department, University of Manchester,
Manchester M13 9PL, United Kingdom.}
\author{David Saad}
\address{Neural Computing Research Group, Aston University, Birmingham
B4 7ET, United Kingdom.}
\date{\today}

\title{Analysis of Natural Gradient Descent for Multilayer Neural Networks}

\maketitle

\def\mxi{\mbox{\boldmath{$\xi$}}}
\def\J{{\bf J}}
\def\B{{\bf B}}
\def\G{{\bf G}}
\def\eg{\mbox{\large $\epsilon$}_{g}}
\def\er{\mbox{\large $\epsilon$}}
\def\tr{{\mathrm T}}
\def\rd{{\mathrm d}}
\def\m{{\mathrm m}}

\begin{abstract}

Natural gradient descent is a principled method for adapting the
parameters of a statistical model on-line using an underlying
Riemannian parameter space to redefine the direction of steepest
descent. The algorithm is examined via methods of statistical
physics which accurately characterize both transient and asymptotic behavior. A
solution of the learning dynamics is obtained for the case of
multilayer neural network training in the limit of large input
dimension. We find that natural gradient learning leads to optimal
asymptotic performance and outperforms gradient descent in the
transient, significantly shortening or even removing plateaus in the
transient generalization performance which typically hamper gradient
descent training.

\end{abstract}
\pacs{87.10.+e, 02.50.-r, 05.20.-y}


\section{Introduction}

One of the most popular forms of neural network training is on-line
learning, in which training examples (input-output pairs) are
presented sequentially and independently at each learning iteration
(for an overview of on-line learning in neural networks,
see~\cite{book}). Natural gradient descent (NGD) was recently proposed by Amari as a
principled alternative to standard on-line gradient descent
(GD)~\cite{amari}. When learning the parameters of a statistical
model, in our case a feedforward neural network, this algorithm has
the desirable properties of asymptotic optimality, given a realizable
learning problem and differentiable model, and invariance to
reparametrizations of our model distribution. NGD is already
established as a popular on-line algorithm for Independent Component
Analysis~\cite{amari} and shows much promise for other statistical
learning problems. Yang and Amari recently introduced an NGD algorithm
for training a multilayer perceptron~\cite{yang,yang2}. In this paper
we provide an analysis of NGD for this problem using a statistical
mechanics formalism. Our results indicate that NGD provides
significantly improved performance over GD and we quantify these gains
for both the transient and asymptotic stages of learning (preliminary
results from this work have been reported in~\cite{rsa}).

The intuition behind NGD comes from viewing the parameter space of a
statistical model as a Riemannian space. A natural measure of
infinitesimal distance between probability distributions is given by
the Kullback-Leibler divergence~\cite{cover}. In this case the Fisher
information matrix can be shown to be the appropriate Riemannian
metric. The natural gradient direction is defined as the direction of
steepest descent under this metric and is obtained by pre-multiplying
the standard Euclidean error gradient with the inverse of the Fisher
information matrix. Since this metric is derived from a divergence
between neighboring model distributions, the algorithm is clearly
independent of model parametrization. An additional beneficial
feature of using this matrix pre-multiplier is that it remains
positive-definite and therefore ensures convergence to a minimum of
the generalization error (assuming the learning rate is annealed
appropriately). This is to be contrasted with other variable-metric
algorithms which utilize the inverse averaged Hessian
matrix. Pre-multiplying the error gradient with the inverse Hessian
may make other fixed points stable, so that the algorithm could
converge to maxima or saddle points on the mean error
surface. Although such methods can be adapted to ensure a
positive-definite matrix pre-multiplier, such adaptations are rather
ad-hoc in nature and are not theoretically well motivated outside of
the asymptotic regime.

Variable-metric methods are often difficult to implement as on-line
algorithms since they require the averaging and inversion of a large
matrix. In the case of NGD we require knowledge of the input
distribution in order to calculate the Fisher information matrix. Yang
and Amari discuss methods for pre-processing training examples in
order to obtain a whitened Gaussian process for the
inputs~\cite{yang2}. If this is possible then, when the input
dimension $N$ is large compared to the number of hidden units $K$,
inversion of the Fisher information for two-layer feedforward
networks requires only $O(N^2)$ operations, providing an efficient and
practical algorithm in many cases. Such a simplification is not
possible for Hessian based methods, because the Hessian involves an
average over input-output pairs. In general it will not be possible to
apply this pre-processing because the input distribution may be far
from Gaussian and difficult to estimate. In this case other on-line
methods will be required in order to approximate the NGD algorithm. We
have recently proposed a method based on a matrix momentum
algorithm~\cite{orr} which allows efficient on-line inversion and
averaging of the Fisher information matrix. This algorithm can be
shown to approximate NGD closely and also provides optimal asymptotic
performance, although at the cost of introducing an extra variable
parameter~\cite{silvia}.

Here, we will consider the idealized situation in which we have the
Fisher information matrix at our disposal. We solve the averaged
dynamics of NGD using a statistical mechanics framework which becomes
exact as $N \rightarrow \infty$ for finite $K$ (see, for example,
\cite{ss,barber,rb,biehl,ws,sr,vicente,lss}). This allows us to
compare performance with standard GD in both the transient and
asymptotic phases of learning, so that we can quantify the advantage
that NGD can be expected to provide. Numerical results for a small
network provide evidence of improved performance. In order to obtain
more generic results we introduce a site-symmetric ansatz for the
special case of a realizable learning scenario, so that we can
efficiently explore a broad range of task complexity and
non-linearity. We show that trapping time in an unstable fixed point
which dominates the training time,the symmetric phase, is
significantly reduced by using NGD and exhibits a slower power law
increase as task complexity grows. We also find that asymptotic
performance is greatly improved, with the generalization performance
of NGD equalling the known universal asymptotics for batch
learning~\cite{am}.

\section{Natural gradient descent}

We consider a probabilistic model $p_\J(\zeta|\mxi)$ for the
distribution of a scalar output $\zeta$ given a vector of inputs $\mxi
\in {\Bbb R}^{N}$ which is parameterized by $\J \in {\Bbb
R}^{KN}$. The Kullback-Leibler divergence provides an appropriate
measure for the distance between distributions~\cite{cover} and for
two nearby points in parameter space we find,
\begin{eqnarray}
	\mbox{KL}\left(p_\J(\zeta,\mxi)\,||\,p_{\J+\rd\J}(\zeta,\mxi)\right)
	& \equiv & \int \!\! \rd \mxi \: p(\mxi) \int \!\! \rd \zeta \:
	p_{\J}(\zeta|\mxi) \log\left(
	\frac{p_\J(\zeta|\mxi)}{p_{\J+\rd\J}(\zeta|\mxi)} \right)
	\nonumber \\ & \simeq & \ \rd\J^\tr \: \G \: \rd\J \ ,
\end{eqnarray}
where $\G$ is the Fisher information matrix,
\begin{equation}
	\G(\J) = \int \!\! \rd \mxi \: p(\mxi) \int \!\! \rd \zeta \:
	p_{\J}(\zeta|\mxi) \: \left(\nabla_{\J}\log
	p_\J(\zeta|\mxi)\right) \left(\nabla_{\J}\log
	p_\J(\zeta|\mxi)\right)^\tr \ .
\end{equation}
This matrix provides a Riemannian metric within the space of model
parameters. We choose the training error
$\er_\J(\zeta,\mxi)\propto-\log p_\J(\zeta|\mxi)$. The direction of
steepest descent within this Riemannian space in terms of expected
error is obtained by pre-multiplying the mean Euclidean error gradient
with $\G^{-1}$~\cite{amari}.

In an on-line learning scheme we draw inputs sequentially
$\mxi^\mu:\mu=1,2,\ldots$ from some distribution $p(\mxi)$ each
labelled according to some stochastic rule
$p_\B(\zeta^\mu|\mxi^\mu)$. The NGD algorithm is defined by a
corresponding sequence of weight updates,
\begin{equation}
	\J^{\mu+1} = \J^\mu + \frac{\eta}{N}\:\G^{-1}\nabla_\J\log
	p_\J(\zeta^\mu|\mxi^\mu) \ .
\label{J_update}
\end{equation}
where the learning rate is scaled by the input dimension for
convenience. This algorithm therefore utilizes an unbiased estimate of
the steepest descent direction in our Riemannian parameter space. If
the rule can be realized by the model and exemplars are corrupted by
output noise then annealing the learning rate
as $\eta=1/\alpha$ at late times (where $\alpha\equiv\mu/N$) results
in optimal asymptotic performance in terms of the quadratic estimation
error, saturating the Cramer-Rao lower bound and equalling in
performance even the best batch algorithms~\cite{amari}.

Consider the deterministic mapping $\phi_\J(\mxi) = \sum_{i=1}^{K}
g\,(\J_{i}^\tr\mxi)$ which defines a soft committee machine (we call
this the student network), where $g(x)$ is some sigmoid activation
function for the hidden units, $\J \equiv \{\J_{i} \}_{1 \leq i \leq
K}$ is the set of input to hidden weights and the hidden to output
weights are set to one. We choose the following Gaussian noise model,
\begin{equation}
	p_{\J}(\zeta|\mxi)
	=\frac{1}{\sqrt{2\pi\sigma_{\m}^2}}\exp\left(\frac{-(\zeta -
	\phi_{\J}(\mxi))^2}{2\sigma_{\m}^2} \right) \ .
\label{pJ}
\end{equation}
The Fisher information matrix for this model distribution is given by
${\bf G} \equiv {\bf A}/\sigma_{\m}^2$ with ${\bf A}$ in block form,
\begin{equation}
	{\bf A}_{ij}= \int \!\! \rd \mxi \: p(\mxi)\:g'({\bf
	J}_i^\tr\mxi)\,g'({\bf J}_j^\tr\mxi)\:\mxi\,\mxi^\tr \ .
\end{equation}
A particularly convenient choice for activation function is $g(x)
\equiv \mbox{erf}(x/\sqrt{2})$ as this allows the average over inputs
to be carried out analytically for an isotropic Gaussian input
distribution $p(\mxi)={\cal N}({\bf 0},{\bf I})$,
\begin{equation}
\label{def_A}
	{\bf A}_{ij}= \frac{2}{\pi\sqrt{\Delta_{ij}}}\left[{\bf I} -
	\frac{1}{\Delta_{ij}}\left( (1+Q_{jj}){\bf J}_i{\bf J}_i^\tr +
	(1+Q_{ii}){\bf J}_j{\bf J}_j^\tr - Q_{ij}({\bf J}_i{\bf
	J}_j^\tr + {\bf J}_j{\bf J}_i^\tr)\right)\right] \ ,
\end{equation}
where $\Delta_{ij} = (1+Q_{ii})(1+Q_{jj})-Q_{ij}^2$ and
$Q_{ij}\equiv\J_i^\tr\J_j$.

\section{Statistical Mechanics Framework}

In order to analyse NGD beyond the asymptotic regime we use a
statistical mechanics description of the learning process which is
exact in the limit of large input dimension $N$ and provides an
accurate model of mean behaviour for realistic values of
$N$~\cite{ss,barber}. We consider the case where outputs are generated
by a teacher network corrupted by Gaussian noise,
\begin{equation}
	p_{\B}(\zeta^\mu|\mxi^\mu)
	=\frac{1}{\sqrt{2\pi\sigma^2}}\exp\left(\frac{-(\zeta^\mu -
	\phi_{\B}(\mxi^\mu))^2}{2\sigma^2} \right) \ ,
\end{equation}
where $\phi_{\B}(\mxi^\mu)=\sum_{n=1}^{M}
g\,(\B_{n}^\tr\mxi^\mu)$. Due to the flexibility of this
mapping~\cite{cybenko} we can represent a variety of learning
scenarios within this framework. The weight update at each iteration
of NGD is then given by,
\begin{equation} 
	\J_i^{\mu+1} = \J_i^\mu +
\frac{\eta}{N}\sum_{j=1}^K\delta_j^\mu{\bf A}_{ij}^{-1}\mxi^\mu \ ,
\end{equation}
where $\delta_i^\mu\equiv
g'(\J_i^\tr\mxi^\mu)[\phi_\B(\mxi^\mu)-\phi_\J(\mxi^\mu)+\rho^\mu]$
and $\rho^\mu$ is zero-mean Gaussian noise of variance
$\sigma^2$. Notice that knowledge of the noise variance is not
required to execute this algorithm since the contributions from the
Fisher information matrix and log-likelihood cancel (recall Eq.~(\ref{J_update})). The model noise
variance is therefore not included as a variable parameter and the
algorithm is well defined even in the deterministic case where
$\sigma_{\m}^2\rightarrow 0$.

The Fisher information matrix can be inverted using the partitioning
method described in~\cite{yang2} (see appendix~\ref{app_inv}); each
block is some additive combination of the identity matrix and outer
products of the student weight vectors,
\begin{equation}
\label{gen_Ainv}
	{\bf A}_{ij}^{-1} = \theta_{ij}{\bf I} +
	\sum_{kl}\Theta_{kl}^{ij}\J_k\J^\tr_l \ .
\end{equation}
where $\theta_{ij}$ are scalars while $\Theta^{ij}$ are $K$
dimensional square matrices. Using the methods described in~\cite{ss}
it is then straightforward to derive equations of motion for a set of
order parameters ${\bf J}_{i}^\tr{\bf J}_{j} \equiv Q_{ij}$, ${\bf
J}_{i}^\tr{\bf B}_{n} \equiv R_{in}$ and ${\bf B}_{n}^\tr{\bf B}_{m}
\equiv T_{nm}$, measuring the various overlaps between student and
teacher vectors. These order parameters are necessary and sufficient
to determine the generalization error $\eg =
\langle\frac{1}{2}(\phi_\J(\mxi) - \phi_\B(\mxi))^2\rangle_\xi$, which
we defined to be the expected error in the absence of noise~\cite{ss}. The equations
of motion are in the form of coupled first order differential
equations for the order parameters with respect to the normalized
number of examples,
\begin{eqnarray}
	\frac{\rd R_{in}}{\rd\alpha} & = & \eta \: f_{in}(R,Q,T)
	\nonumber \ , \\ \frac{\rd Q_{ik}}{\rd\alpha} & = & \eta \:
	g_{ik}(R,Q,T) + \eta^2 \: h_{ik}(R,Q,T,\sigma^2) \ .
\label{eq_dynamics}
\end{eqnarray}
where $R=[R_{in}]$, $Q=[Q_{ik}]$ and $T=[T_{nm}]$. The explicit
expressions are given in appendix~\ref{app_eom}. These equations can
be integrated numerically in order to determine the evolution of the
generalization error.

\section{Numerical Results}
\label{sec_num}

\begin{figure}[t]
	\begin{center} \leavevmode \epsfysize = 10cm \epsfbox[150 180
	450 480]{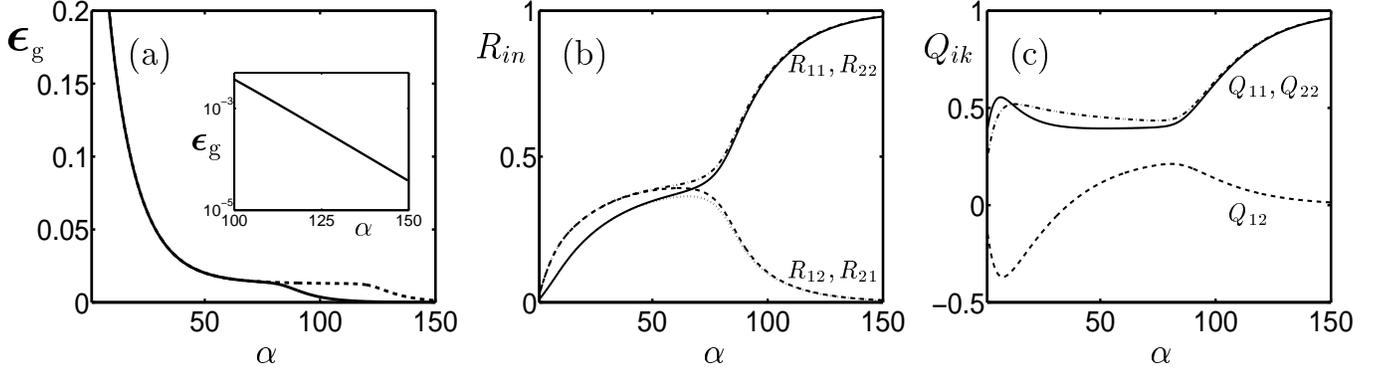} \end{center} \caption{Numerical
	integration of the NGD equations of motion. A two-node soft
	committee machine learns from examples
	generated by a two-node isotropic teacher
	($K=M=2$, $T_{nm}=\delta_{nm}$) in the absence of
	noise. The learning rate is fixed at
	$\eta=0.05$ and initial conditions are $R_{in},Q_{i\neq k}\in
	U[0,10^{-3}]$ and $Q_{ii}\in U[0,0.5]$. The generalization
	error is shown by the solid line in (a) with the exponential asymptotic
	decay shown on a log scale in the inset (the dashed line shows
	the effect of reducing the initial conditions by a factor of
	$10^{-3}$). The student-teacher and student-student overlaps
	are shown in (b) and (c) respectively. \vspace{-2cm}}
	\label{fig_dynamics}
\end{figure}

In Fig.~\ref{fig_dynamics} we show an example of the NGD dynamics for
a realizable and noiseless learning scenario ($K=M=2$,
$T_{nm}=\delta_{nm}$). Fig.~\ref{fig_dynamics}(a) shows the evolution
of the generalization error while Figs.~\ref{fig_dynamics}(b) and (c)
show the student-teacher and student-student overlaps respectively
(the indices have been re-ordered {\em a posteriori}). We have used
initial conditions corresponding to an input dimension of about
$N\simeq10^{6}$, although we expect the dynamical equations to
describe mean behavior accurately for much smaller systems as was
found to be the case for GD~\cite{barber}. The dashed line in
Fig.~\ref{fig_dynamics}(a) shows the effect of reducing the initial
conditions for each $R_{in}$ and $Q_{i\neq k}$ by a factor of
$10^{-3}$, which corresponds to an input dimension of about
$N\simeq10^{12}$. The symmetric phase seems to grow logarithmically
as $N$ increases, as was also found to be the case for
GD~\cite{biehl}.

As is the case for GD~\cite{ss} the dynamics for this example can be
characterized by two major phases of learning, the symmetric phase and
asymptotic convergence. Following a short initial transient the order
parameters are trapped in a subspace
characterized by a lack of differentiation between the activities of
different teacher nodes. After an initial reduction, the
generalization error remains at a constant non-zero value and the
student-teacher overlaps are virtually indistinguishable. This
symmetric phase is an unstable fixed point of the dynamics and
eventually small perturbations due to the random initial conditions
lead to escape and convergence towards zero generalization error. If
the teacher is deterministic, as in this example, then the
generalization error converges to zero exponentially unless the
learning rate is chosen too large (see the inset to
Fig.~\ref{fig_dynamics}(a)). If the teacher's output is corrupted by
noise then the learning rate must be annealed in order for the
generalization error to decay asymptotically (we will consider this
regime in more detail in section~\ref{sec_asymptotic}).

The dynamics differs from the GD result in that the symmetric phase is
typically less pronounced, although the dashed line in
Fig.~\ref{fig_dynamics}(a) shows how the symmetric phase increases in
duration as $N$ increases (because of a reduced asymmetry in the
initial conditions). The dynamics for GD and NGD are qualitatively
different for small learning rates, where fluctuations in the gradient
are completely suppressed and the $\eta^2$ terms in
Eq.~(\ref{eq_dynamics}) can be neglected. In this limit the
symmetric phase disappears completely for NGD, while it still
dominates the learning time for GD. The symmetric phase is a
fluctuation driven phenomena for NGD, rather than a perturbation
around the deterministic result. As described in the next section,
this makes analysis of the symmetric phase more difficult than for GD
since a small learning rate expansion is no longer meaningful.

A quantitative comparison of GD and NGD is difficult because both
algorithms have a free parameter, the learning rate $\eta$, which can
be chosen arbitrarily and which will be critical to performance. In
order to make a principled comparison we choose to
compare the algorithms when their learning rates are chosen to be
optimal. This can be achieved by using a variational method
which allows us to determine the globally optimal time-dependent
learning rate for each algorithm~\cite{sr}. The resulting learning
rate optimizes the total change in generalization error over a fixed
time-window and is found by extremitization of the following functional
(see~\cite{sr} for details and results for optimized GD):
\begin{equation}
	\Delta \mbox{\large$\epsilon$}_{g} [\eta(\alpha)] \ = \
	\int_{\alpha_0}^{\alpha_1} \frac{\rd
	\mbox{\large$\epsilon$}_{g}}{\rd\alpha} \ \rd\alpha \ = \
	\int_{\alpha_0}^{\alpha_1} \cal{L}[\eta(\alpha),\alpha] \
	\rd\alpha \ .
\label{eq_opt}
\end{equation}
Numerical results suggest that the optimal learning rates determined
here are close to the critical learning rate within the symmetric
phase for both methods, above which the student weight vector norms
increase without bound.

\begin{figure}[t]
	\begin{center} \leavevmode \epsfysize = 10cm \epsfbox[150 180
	450 480]{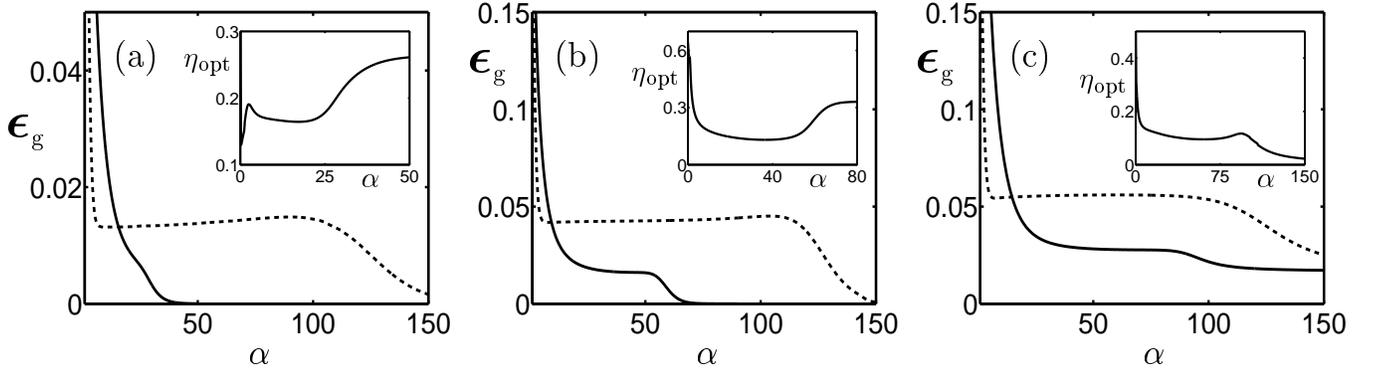} \end{center} \caption{The optimal
	performance of a two node student ($K=2$) for NGD (solid
	line) and GD (dashed line) for (a) over-realizable: $M=1$, (b)
	realizable: $M=2$ and (c) unrealizable: $M=3$ learning
	scenarios. The optimal learning rate schedule for NGD is shown
	by the inset to each figure. The teacher is isotropic
	($T_{nm}=\delta_{nm}$), noise free and initial conditions are
	as in Fig.~\ref{fig_dynamics}}. 
	\label{fig_opt}
\end{figure}

In Fig.~\ref{fig_opt} we compare the performance of optimized GD and
optimized NGD for a two-node student learning from a noiseless,
isotropic teacher starting from the same initial conditions ($K=2$,
$T_{nm}=\delta_{nm}$). Figs.~\ref{fig_opt}(a), (b) and (c) show
results for teachers with $M=1$, $M=2$ and $M=3$ hidden nodes
respectively. In each case the optimal learning rate schedule for NGD
is shown by the inset. It should be noted that although there is a
significant temporal variation in the optimized $\eta$, very similar
performance would be achieved by choosing $\eta$ to be fixed at its
average value. We see that NGD significantly outperforms GD in each
example. For the over-realizable example shown in
Fig.~\ref{fig_opt}(a) the difference is most significant, with NGD
displaying no obvious symmetric plateau. Performance of the NGD
algorithm seems to reflect the difficulty inherent in the task, while
GD displays very similar performance in each case. It is interesting
to compare our results with those found using a locally optimal rule
derived by variational arguments~\cite{vicente}. The variational
approach requires rather detailed information about the teacher's
structure and would be difficult to approximate with a practical
algorithm. However, we find rather similar performance with NGD,
especially for the $K=M=2$ example shown both here and
in~\cite{vicente}. The performance bottleneck for GD is due to an
inherent symmetry in the student parametrization, while for NGD the
task complexity seems to be more important. Also notice that the
generalization error is significantly lower during the symmetric
plateau for NGD in each case, which is due to reduced weight vector
norms (this is also true for the locally optimal algorithm). It is the
growth of these norms which limits increases in the learning rate for
GD and it appears that NGD is much more effective in controlling this
effect. Another interesting difference between the NGD and GD dynamics is in
the short transient prior to the symmetric phase. The NGD dynamic
seems to converge much slower to the symmetric fixed point, as shown
in Fig.~\ref{fig_opt}, reflecting the fact that the strong
eigenvalues, related to eigenvectors which lead the dynamic to the
symmetric pixed point, are effectively reweighed and suppressed by the
NGD rule.

\section{Generic Results for a Symmetric System}

Although our equations of motion are sufficient to describe learning
for arbitrary system size, the number of order parameters is
$\frac{1}{2}K(K+1)+KM$ so that the numerical integration soon becomes
rather cumbersome as $K$ and $M$ grow and analysis becomes
difficult. To obtain generic results in terms of system size we
therefore exploit symmetries which appear in the dynamics for
isotropic tasks and structurally matched student and teacher ($K=M$
and $T=T\delta_{nm}$). This site-symmetric ansatz is only rigorously
justified for the special case of symmetric initial conditions and
further investigations are required to determine the validity of this
approximation in general for large values of $K$ (fixed points other
than those considered here have been reported for GD~\cite{biehl} and
it is unclear whether or not their basins of attraction are
negligible).  Simulations of the GD dynamics for $K$ up to $10$, with
random initial conditions, show good correspondence with the symmetric
system. In this case we define a four dimensional system via
$Q_{ij}=Q\delta_{ij}+C(1-\delta_{ij})$ and
$R_{in}=R\delta_{in}+S(1-\delta_{in})$ which can be used to study the
dynamics for arbitrary $K$ and $T$ (here, $\delta_{ij}$ denotes the
Kronecker delta). In appendix~\ref{app_inv} we show how the Fisher
information matrix can be inverted for the reduced dimensionality
system and the resulting equations of motion are given in
appendix~\ref{app_eom_red}.

Analytical study of the symmetric phase for GD is only feasible for
small learning rates, since in this case the symmetric fixed point is
easily determined and a linear expansion around this fixed point is
possible~\cite{ss,ws}. Such an analysis is not feasible for NGD
because the dynamics never approaches this fixed point (the Fisher
information matrix becomes singular when $Q=C$). In any case, a small
$\eta$ analysis will be of limited value since it is the fluctuation
driven terms in the dynamics (terms proportional to $\eta^2$ in
Eq.~(\ref{eq_dynamics})) which set the learning
time-scale and determine the optimal and maximal learning rate during
the symmetric phase. In order to study the performance of both methods
for larger learning rates we will therefore apply a the optimal
learning rate framework described in the preceding section~\cite{sr}.

The impact of output noise on the symmetric phase dynamics is not
considered explicitly here. For low noise levels there is no
noticeable effect on the length of the symmetric phase, or on the
order parameters and generalization error within this phase. For
larger noise levels the symmetric phase increases in length and the
student norms increase, resulting in a larger generalization
error. We expect that these are secondary effects and that most
essential features of this phase are captured by the noiseless
dynamics. This is not true for later stages of learning, where the
inclusion of noise completely alters qualitative features of the
dynamics. These asymptotic effects are considered in
section~\ref{sec_asymptotic} below.

\subsection{Globally optimal performance}

The optimal learning rate is determined as describe before
Eq.~(\ref{eq_opt}) in section~\ref{sec_num}. In the following
examples we use a brief initial learning phase with GD (until
$\alpha=1$) as this results in faster entry into the symmetric phase
and also leads to quicker convergence of the learning rate
optimization. The effect on learning time will be negligible as $K$
becomes very large, but this procedure might be used to improve
performance in practice for realistically sized networks.

\begin{figure}[t]
	\begin{center} \leavevmode \epsfysize = 10cm \epsfbox[150 140
	450 440]{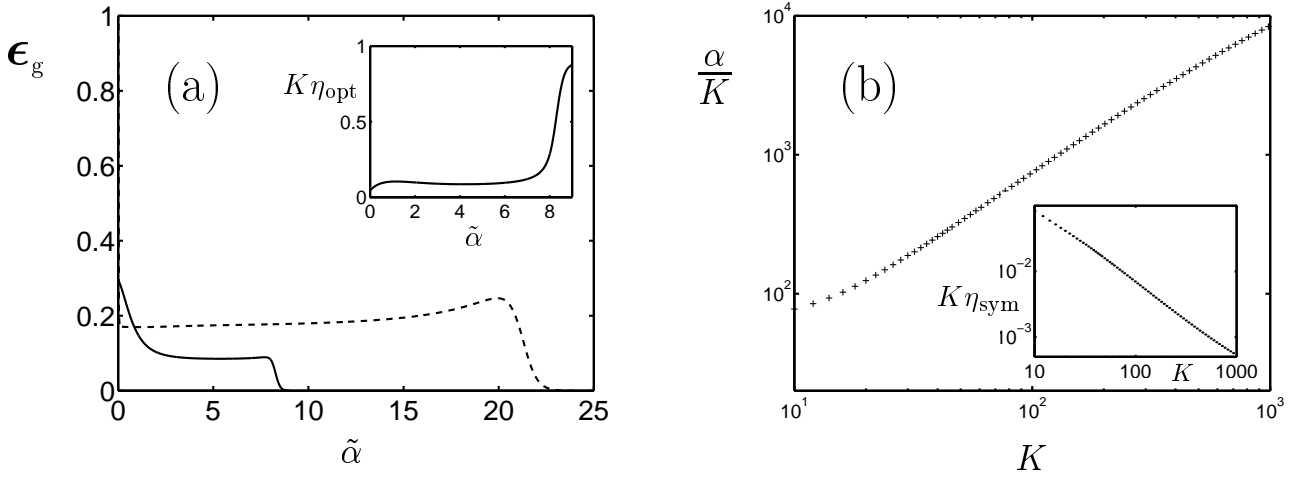} \end{center} \caption{In (a) the
	generalization error is shown for optimal
	NGD (solid line) and optimal gradient
	descent (dashed line) for $K=10$ in the site-symmetric system (we define
	$\tilde{\alpha}=10^{-2}\alpha$). The inset shows the optimal
	learning rate for NGD. In (b) the time required for
 	optimal NGD to reach a generalization error of $10^{-4}K$ is
	shown as a function of $K$ on a log-log scale. The inset shows
	the optimal learning rate within the symmetric phase. In both
	(a) and (b) we used $T=1$, zero noise and initial conditions
	$R=10^{-3}$, $Q = U[0,0.5]$ and $S=C=0$. A brief stage of GD
	is used before NGD is started.}
	\label{fig_symphase}
\end{figure}

Fig.~\ref{fig_symphase} summarises our results for transient
learning in the absence of noise. In Fig.~\ref{fig_symphase}(a) we
compare optimal performance for $K=10$ and $T=1$, which indicates a
significant shortening of the symmetric phase for NGD (the inset shows
the optimal learning rate for NGD). Fig.~\ref{fig_symphase}(b) shows
the time required for NGD to reach a generalization error of
$10^{-4}K$ as a function of $K$ (for $T=1$). The learning time is
dominated by the symmetric phase, so that these results provide a
scaling law for the length of the symmetric phase in terms of task
complexity. We find that the escape time for NGD scales as $K^2$,
while the inset shows that the learning rate within the symmetric
phase approaches a $K^{-2}$ decay. Scaling laws for GD were determined
in~\cite{ws} (also using a site-symmetric ansatz), showing a
$K^{\frac{8}{3}}$ law for escape time and a learning rate scaling of
$K^{-\frac{5}{3}}$ within the symmetric phase. The escape time for the
adaptive learning rule studied in~\cite{ws} scales as
$K^{\frac{5}{2}}$, which is also worse than NGD.

\subsection{Asymptotic convergence with noise}
\label{sec_asymptotic}

\begin{figure}[t]
	\begin{center} \leavevmode \epsfysize = 10cm \epsfbox[150 180
	450 480]{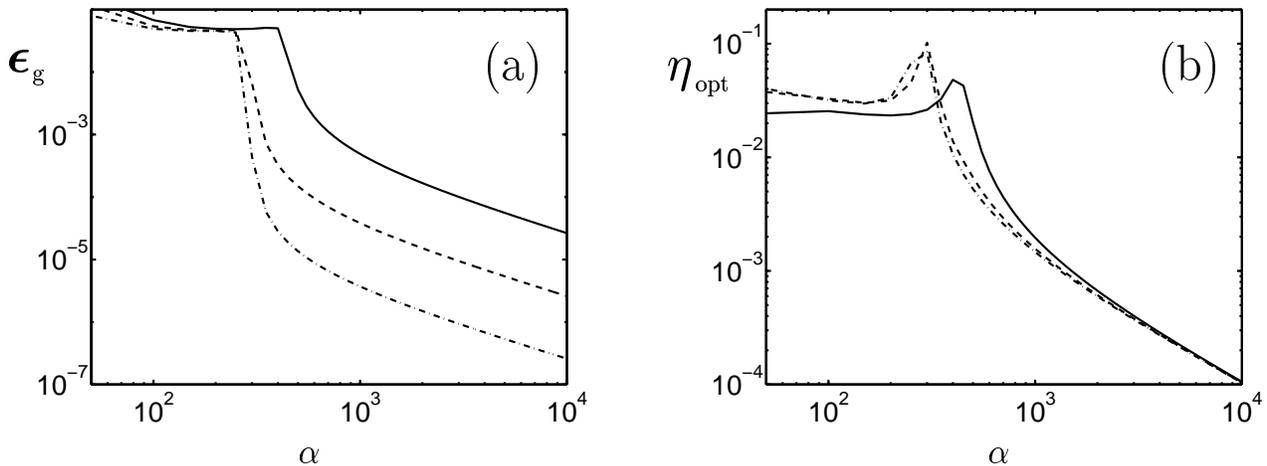} \end{center} \caption{Optimal NGD with
	teacher corrupted by Gaussian noise for
	$K=5$ and $T=1$ in the site-symmetric system, shown on a
	log-log scale. The
	generalization error and optimal time-dependent learning rates
	are shown in (a) and (b) respectively, with $\sigma^2=0.1$
	(solid line), $\sigma^2=0.01$ (dashed line) and
	$\sigma^2=0.001$ (dot-dashed line).}  \label{fig_noise}
\end{figure}

After the symmetric phase, the order parameters begin convergence
towards their asymptotic values ($R_\infty=Q_\infty=T$,
$S_\infty=C_\infty=0$) and for the realizable scenario considered here
the generalization error converges towards zero (recall that we have
defined the generalization error to be the expected error in the
absence of noise). In the absence of output noise this convergence is
exponential for a fixed learning rate so long as we do not choose the
learning rate too high. However, in the presence of output noise the
learning rate must be annealed in order to achieve zero generalization
error asymptotically. It is known that NGD is asymptotically optimal,
in terms of the covariances of the student-teacher weight deviations
(the quadratic estimation error), with $\eta=1/\alpha$, saturating the
Cramer-Rao bound and equalling in performance even the best batch
methods~\cite{amari}. However, the quadratic estimation error has no
direct interpretation in terms of generalization ability. In
Fig.~\ref{fig_noise} we show results for optimized NGD dynamics with
$K=5$ and $T=1$. Fig.~\ref{fig_noise}(a) shows the generalization
error and Fig.~\ref{fig_noise}(b) shows the corresponding optimal
learning rate schedules for three noise levels ($\sigma^2 = 0.1$,
$\sigma^2 = 0.01$ and $\sigma^2 = 0.001$). The graphs are on log-log
scales and show that the optimized learning rates indeed converge to
a $1/\alpha$ decay after leaving the symmetric phase. The
generalization error decays at the same rate, but with a prefactor
which depends on the noise level.

In order to determine the asymptotic generalization error decay
analytically we apply recent results for the annealing dynamics of
GD~\cite{lss}. This allows a comparison between the asymptotic
generalization error for NGD and the result for GD. In
appendix~\ref{app_eom_asy} we solve the asymptotic dynamics for
annealed learning. As expected, the optimal annealing schedule for NGD
is found by setting $\eta=1/\alpha$ at late times. By contrast,
although the optimal learning rate for GD is also inversely
proportional to $\alpha$, the optimal prefactor depends on $K$ and $T$
in a non-trivial manner~\cite{lss}. For both optimized GD and NGD the
generalization error decays according to an inverse power law:
\begin{equation}
	\mbox{\large $\epsilon$}_{g} \sim \frac{\mbox{\large
	$\epsilon$}_{0}\sigma^2}{\alpha} \quad \mbox{ as } \: \:
	\alpha \rightarrow \infty \ .
\end{equation}
The exact result for NGD takes a very simple form $\mbox{\large
$\epsilon$}_{0} = \mbox{$\frac{1}{2}$}K$ independent of the value of
$T$. This equals the universal asymptotics for optimal maximum
likelihood and Bayes estimators which depend only on the learning
machine's number of degrees of freedom~\cite{am}. NGD is therefore
asymptotically optimal in terms of both generalization error and
quadratic estimation error.

\begin{figure}[t]
	\begin{center} \leavevmode \epsfysize = 10cm \epsfbox[150 140
	450 440]{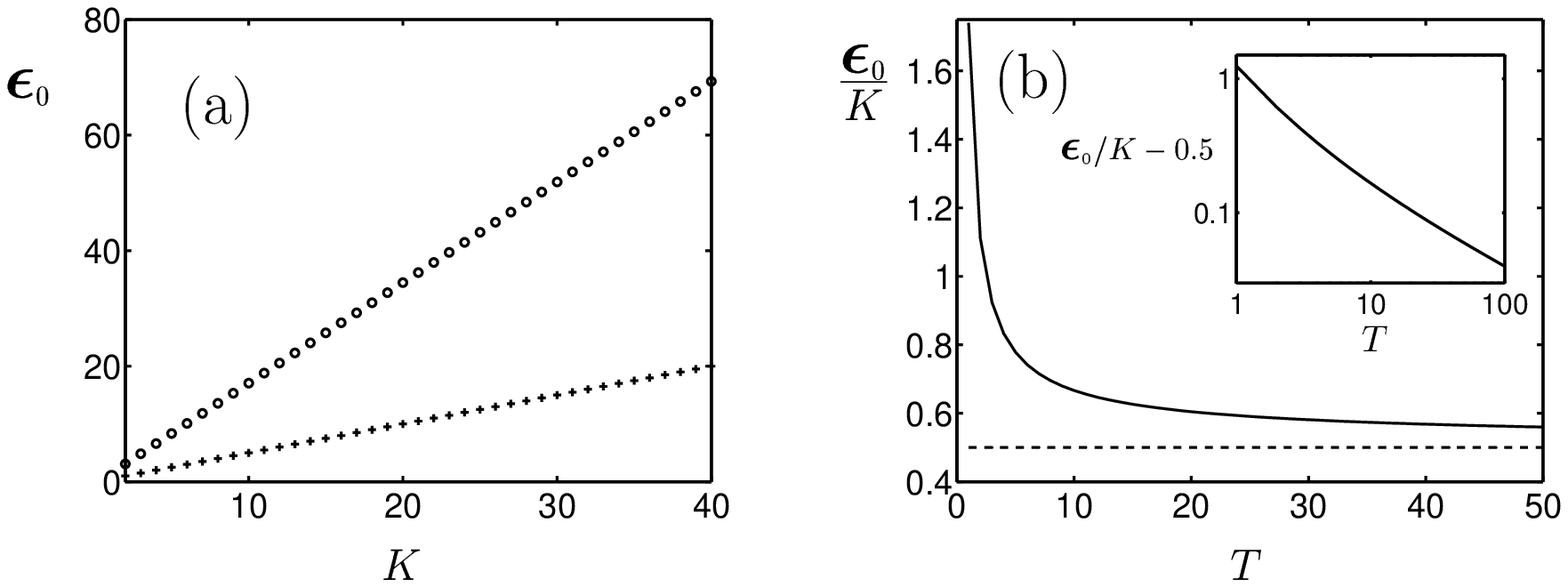} \end{center} \caption{Prefactor for
	the asymptotic decay of the generalization error: (a) shows
	the prefactor for $T=1$ as a function of $K$ for optimal GD
	(circles) and NGD (crosses) while (b)
	shows how the prefactor for optimal GD (large $K$)
	decays towards $K/2$ as $T$ increases, which is the
	prefactor for NGD. The inset to (b) shows the GD result on a
	log-log scale.}  \label{fig_asympt}
\end{figure}

In Fig.~\ref{fig_asympt} we compare the prefactor of the
generalization error decay for NGD and optimal
GD. Fig.~\ref{fig_asympt}(a) shows the result for $T=1$ as a function
of $K$, indicating an approximately linear scaling law for GD (The
result above shows that the NGD scaling is linear in $K$). In
Fig.~\ref{fig_asympt}(b) we compare the decay prefactors for each
method as a function of $T$, showing how the difference diverges as
$T$ is reduced (the GD results are for large $K$). This can be
explained by examining the asymptotic expression for the Fisher
information matrix, shown in Eq.~(\ref{asy_A}). For large $T$ the
diagonals of this matrix are $O(1/\sqrt{T})$ and equal (for large $N$)
while all other terms are at most $O(1/T)$, so that the Fisher
information is effectively proportional to the identity matrix in this
limit and NGD is asymptotically equivalent to GD. However, for small
$T$ the diagonals are $O(T^2)$ while the off-diagonals remain finite,
so that the Fisher information is dominated by off-diagonals in this
limit.

\section{Conclusion}

We have used a statistical mechanics formalism to solve and analyse
the dynamics of natural gradient descent (NGD) for learning in a
two-layer feedforward neural network. In order to quantify the
comparative performance of NGD and gradient descent (GD) we compared
the optimized performance of each algorithm by determining the optimal
learning rate in each case. We found that NGD provided significant
gains in performance over GD in every case examined, both in the
transient and asymptotic stages of learning. A site-symmetric ansatz
was applied in order to simplify the dynamical equations for a
realizable and isotropic task. This allowed the dynamics of large
networks to be integrated efficiently so that we could determine
generic behaviour for large networks. We found that the learning time
scaled as $K^2$ where $K$ is the number of hidden nodes, compared to a
scaling of $K^{\frac{8}{3}}$ for GD~\cite{ws}. Asymptotically NGD is
known to provide optimal performance with $\eta=1/\alpha$ in terms of
the quadratic estimation error. An asymptotic solution to the annealed
learning rate dynamics showed this schedule to also be optimal in
terms of generalization error, with the error decay saturating the
universal asymptotics for optimal maximum likelihood and Bayes
estimators~\cite{am}. We compared this result with the optimized
schedule for GD and plotted the relative performance for various
values of task-nonlinearity $T$. The difference in performance was
found to be largest for small values of $T$. However, in the case of
NGD the optimal annealing schedule at late times is known, while for
GD it is a complex function of $K$ and $T$ which will be difficult to
estimate in general.

One possible drawback for NGD is the rather complex transient
behavior of the optimal learning rate. For example, in the realizable
isotropic case the optimal learning rate scales as $K^{-2}$ in the
symmetric phase and $K^{-1}$ asymptotically in the absence of
noise. It is also unclear where learning rate annealing should begin
in the presence of output noise. Asymptotically the optimal annealing
schedule is known, so the situation is better than for GD, but the
problem of setting a good learning rate in the transient remains. In
practical applications there will also be an increased cost required
in estimating and inverting the Fisher information
matrix~\cite{yang2}. Here, we have only considered the idealized
situation in which the Fisher information matrix is exactly
known. In~\cite{silvia} we adapt a matrix momentum algorithm due to
Orr and Leen~\cite{orr} in order to obtain efficient averaging and
inversion of the Fisher information matrix on-line. This algorithm is
shown to provide a good approximation to NGD, although this is at the
cost of including an extra parameter.

\section*{Acknowledgements} We would like to thank Shun-ichi Amari for
useful discussions. This work was supported by the EPSRC grant
GR/L19232.

\appendix

\section{Inverting the Fisher information matrix}
\label{app_inv}

In general the Fisher information matrix should be inverted using the
block inversion method described in~\cite{yang2}. The parameters in
Eq.~(\ref{gen_Ainv}) are then complicated functions of $Q$ which must
be determined iteratively (see~\cite{newton} for a similar method
applied to the Hessian matrix for $M=K=2$). Below we consider the simpler
situation of a site-symmetric system, in which case the inversion can
be carried out in closed form for arbitrary $K$. Asymptotically the
result is shown to be further simplified.

\subsection{Site-symmetric system}

Exploiting symmetries in the dynamics of realizable isotropic learning
($K=M$ and $T_{nm}=T\delta_{nm}$) we consider a reduced dimensionality
system with $Q_{ij} = Q\delta_{ij} + C(1-\delta_{ij})$. We can then
write block $(i,j)$ of the Fisher information matrix as (see
Eq.~(\ref{def_A})),
\begin{equation} 
	{\bf A}_{ij} = (a\:\delta_{ij}+b)\:{\bf I} +
	(c\:\delta_{ij}+d)\:{\bf J}_i{\bf J}_i^\tr + d\:{\bf J}_j{\bf
	J}_j^\tr + e\:({\bf J}_i{\bf J}_j^\tr + {\bf J}_j{\bf
	J}_i^\tr) \ ,
\label{def_Asym}
\end{equation}
where we have defined,
\begin{eqnarray*}
	& & b = \frac{2}{\pi\sqrt{((1+Q)^2-C^2)}} \ , \quad a =
	\frac{2}{\pi\sqrt{1+2Q}} - b \ , \quad c =
	\frac{4\,(1+Q-C)}{\pi((1+Q)^2-C^2)^{\frac{3}{2}}} -
	\frac{4}{\pi(1+2Q)^{\frac{3}{2}}}\ , \\ & & d =
	\frac{-2\,(1+Q)}{\pi((1+Q)^2-C^2)^{\frac{3}{2}}} \ , \quad e =
	\frac{2C}{\pi((1+Q)^2-C^2)^{\frac{3}{2}}} \ .
\end{eqnarray*}
Block $(i,j)$ in the inverse of ${\bf A}$ is then given by,
\begin{equation}
	{\bf A}^{-1}_{ij} = \left(\frac{1}{a}\:\delta_{ij} -
	\frac{b}{a(a+bK)}\right)\,{\bf I} \ + \
	\sum_{k=1}^K\sum_{l=1}^K \Gamma^{kl}_{ij}{\bf J}_k{\bf
	J}_l^\tr \ ,
\label{def_Ainv}
\end{equation}
and symmetries suggests the following general form for ${\bf \Gamma}$,
\begin{eqnarray}
	& & \hspace{-1cm} \Gamma_{ij}^{kl} = \gamma_1
	\delta_{ij}\delta_{ik}\delta_{il} +
	\gamma_2(\delta_{ik}\delta_{il}+\delta_{jk}\delta_{jl}) +
	\gamma_3(\delta_{ik}\delta_{jk}+\delta_{il}\delta_{ij}) +
	\gamma_4\delta_{kl}\delta_{ij} +
	\gamma_5\delta_{ik}\delta_{jl} \nonumber \\ & & + \
	\gamma_6\delta_{jk}\delta_{il} +
	\gamma_7(\delta_{jl}+\delta_{ik}) +
	\gamma_8(\delta_{jk}+\delta_{il})+\gamma_9\delta_{kl} +
	\gamma_{10}\delta_{ij} + \gamma_{11} \ .
\end{eqnarray}
We therefore have to set $11$ free parameters in order to fully
specify ${\bf \Gamma}$. This is achieved by substituting
Eqs.~(\ref{def_Asym}) and (\ref{def_Ainv}) into the definition of
the inverse,
\begin{equation}
	\sum_{k=1}^K {\bf A}_{ik} {\bf A}^{-1}_{kj} \ = \
	\delta_{ij}{\bf I} \quad \forall \: i,j \ .
\label{def_invA}
\end{equation}
Equating like terms leads to a set of $15$ equations and we can choose
any linearly independent subset of $11$ equations in order to
determine {\boldmath $\gamma$}. For one particular choice we find,
\begin{equation}
	\mbox{\boldmath $\gamma$} = {\bf M}^{-1}\mbox{\boldmath
	$\beta$} \ ,
\label{def_gam}
\end{equation}
where the non-zero terms in ${\bf M}_{11\times 11} = [m_{i,j}]$ and
$\mbox{\boldmath $\beta$}_{1\times 11} = [\beta_i]$ are defined below,
\begin{eqnarray*}
	& & m_{1,1} = m_{2,2} = m_{4,3} = (Q-C)(d+e) + c\,Q + a \ , \\
	& & m_{1,2} = m_{2,9} = m_{4,8} = c\,[Q+C(K-1)] \ , \\ & &
	m_{1,3} = m_{2,8} = m_{4,10} = (Q-C)(dK+c) \ , \\ & & m_{1,4}
	= m_{1,6} = c\,(Q-C) \ , \quad m_{2,4} = m_{4,6} = c\,C \\ & &
	m_{2,6} = m_{4,4} = m_{5,4} = m_{5,6} = d(Q-C) \ , \\ & &
	m_{3,2} = m_{6,4} = m_{8,6} = m_{11,8} = a \ , \\ & & m_{3,3}
	= m_{6,6} = m_{7,4} = m_{7,6} = m_{8,4} = m_{11,10} = e(Q-C) \
	, \\ & & m_{5,1} = m_{9,5} = b + dQ + eC \ , \\ & & m_{5,2} =
	m_{7,2} = m_{10,9} = (d+e)(Q+C(K-1)) \ , \\ & & m_{5,3} =
	d(Q-C) + Kb + a \ , \quad m_{7,1} = m_{10,2} = m_{10,3} =
	eQ+dC \ , \\ & & m_{7,3} = m_{10,10} = e(Q-C) + c\,C + (d+e)KC
	\ , \\ & & m_{7,5} = m_{10,7} = (Q-C)(d+c+e)+a+c\,C+K(Qe+Cd) \
	, \\ & & m_{7,7} = m_{10,11} = (Q-C)(K(d+e)+c) + c\,CK +
	(d+e)CK^2 \ , \\ & & m_{9,2} = b \ , \quad m_{9,3} = m_{10,4}
	= m_{10,6} = (d+e)C \ , \\ & & m_{9,7} = Kb+a+(d+e)(Q+C(K-1))
	\ , \\ & & m_{10,8} = (Q-C)(d+2e)+c\,C+2KC(d+e) \ , \\ \\ & &
	\beta_1 = -\frac{c}{a} \ , \quad \beta_4 =
	\frac{b(c+dK)}{a(a+bK)} - \frac{d}{a}\ , \quad \beta_5 =
	-\frac{d}{a} \\ & & \beta_7 = \beta_8 = -\frac{e}{a} \ , \quad
	\beta_{10} = \beta_{11} = \frac{eb}{a(a+bK)} \ .
\end{eqnarray*}
 
\subsection{Asymptotic inversion}
\label{app_inv_asy}

For realizable rules the asymptotic form for each block of {\bf A} is
(to leading order),
\begin{equation}
	{\bf A}_{ij} = (a\,\delta_{ij}+b)\:{\bf I} +
	(c\,\delta_{ij}+d)\:{\bf B}_i{\bf B}_i^\tr + d\:{\bf B}_j{\bf
	B}_j^\tr \ ,
\label{asy_A}
\end{equation}
where,
\begin{eqnarray*}
	& & a = \frac{2}{\pi\sqrt{1+2T}} - \frac{2}{\pi(1+T)} \ ,
	\quad b = \frac{2}{\pi(1+T)} \ ,\\ & & c =
	\frac{4}{\pi(1+T)^2} - \frac{4}{\pi(1+2T)^{\frac{3}{2}}} \ ,
	\quad d = -\frac{2}{\pi(1+T)^2} \ .
\end{eqnarray*}
Block $(i,j)$ in the inverse of ${\bf A}$ is then given by,
\begin{equation}
	{\bf A}^{-1}_{ij} = \left(\frac{1}{a}\:\delta_{ij} -
	\frac{b}{a(a+bK)}\right)\,{\bf I} \ + \ \sum_{n=1}^K
	\Gamma^{n}_{ij}{\bf B}_n{\bf B}_n^\tr \ ,
\label{asy_Ainv}
\end{equation}
Substituting these expressions into Eq.~(\ref{def_invA}) and
using the orthogonality of the teacher weight vectors
($T_{nm}=T\delta_{nm}$) we obtain a matrix equation for ${\bf
\Gamma}^n = [\Gamma_{ij}^n]$,
\begin{equation}
	{\bf \Gamma}^n = {\bf P}^{-1}{\bf X}
\end{equation}
where,
\begin{eqnarray*}
	{\bf P} & = & a \, {\bf I} + \left(\begin{array}{c} {\bf e}_n
	\\ {\bf u} \end{array}\right)^\tr \left(\begin{array}{c c} cT
	& -dT \\ -dT & b \end{array}\right) \left(\begin{array}{c}
	{\bf e}_n \\ {\bf u} \end{array}\right) \ , \\ {\bf X} & = &
	\frac{1}{a}\left(\begin{array}{c} {\bf e}_n \\ {\bf u}
	\end{array}\right)^\tr \left(\begin{array}{c c} -c &
	(da-bc)/(a+bK) \\ d & -db/(a+bK) \end{array}\right)
	\left(\begin{array}{c} {\bf e}_n \\ {\bf u} \end{array}\right)
	\ .
\end{eqnarray*}
Here, we have defined ${\bf e}_n$ to be a $K$-dimensional row vector
with a one in the $n$th element and zeros everywhere else, while ${\bf
u}$ is a row vector of ones. Solving for ${\bf \Gamma}^n$ we find,
\begin{equation}
	{\bf \Gamma}^n = \frac{1}{\Delta}\left(\begin{array}{c} {\bf
	e}_n \\ {\bf u} \end{array}\right)^\tr {\bf \Theta}
	\left(\begin{array}{c} {\bf e}_n \\ {\bf u} \end{array}\right)
	\ ,
\end{equation}
where,
\begin{eqnarray*}
	{\bf \Theta} & = & \left(\begin{array}{c c} K(d^2T-bc)-ac &
	d^2T-bc-ad \\ d^2T-bc-ad & (d^2T(a+b) - 2abd - b^2c)/(a+bK)
	\end{array}\right) \ , \\ \Delta & = &
	a^2(a+bK)+a^2T(c-2d)+aT(K-1)(cb-d^2) \ .
\end{eqnarray*}

\section{Equations of motion}
\label{app_eom}

Using the definition of ${\bf A}^{-1}$ given in
Eq.~(\ref{gen_Ainv}) we find,
\begin{eqnarray}
  & & \frac{\rd R_{in}}{\rd\alpha} = \eta\sum_j\left(
\theta_{ij}\phi_{jn} + \sum_{kl}\Theta_{kl}^{ij} R_{kn}
\psi_{jl}\right) \ , \nonumber \\ & & \frac{\rd Q_{ir}}{\rd\alpha} =
\eta\sum_j\left( \theta_{ij}\psi_{jr} + \theta_{rj}\psi_{ji} +
\sum_{kl} \psi_{jl}( \Theta_{kl}^{ij} Q_{kr} + \Theta_{kl}^{rj}
Q_{ki})\right) + \eta^2 \sum_{jk} \theta_{ij}\theta_{rk}\upsilon_{jk}
\ .
\label{newton_eom}
\end{eqnarray}
Here $\phi_{in} \equiv \langle\delta_{i} y_{n} \rangle_{\{ \xi \}}$,
$\psi_{ik} \equiv \langle\delta_{i} x_{k}\rangle_{\{ \xi \}}$ and
$\upsilon_{ik} \equiv \langle \delta_{i} \delta_{k} \rangle _{\{ \xi
\}}$ where $x_i\equiv\J_i^\tr\mxi$ and $y_n\equiv\B_n^\tr\mxi$ are
activations of the $i$th student and
$n$th teacher hidden nodes respectively and \mbox{$\delta_{i}^{\mu} \equiv
g'(x^{\mu}_{i})\bigl[\sum_{n=1}^{M} g(y^{\mu}_{n}) - \sum_{j=1}^{K}
g(x^{\mu}_{j}) + \rho^\mu\bigr]$}. The brackets denote averages over
inputs which can be written as averages over the multivariate Gaussian distribution
of student and teacher activations. The explicit expressions for $\phi_{in}$, $\psi_{ik}$,
$\upsilon_{ik}$ depend exclusively on the weight overlaps (the
covariances of the activation distribution) and are given in~\cite{ss}.

\subsection{Site-symmetric system}
\label{app_eom_red}

We substitute the definition of the inverse Fisher information for a
symmetric system from Eq.~(\ref{def_Ainv}) into
Eq.~(\ref{J_update}) to get the weight update equation:
\begin{equation}
	{\bf J}_i^{\mu+1} = {\bf J}_i^\mu +
	\frac{\eta}{N}\Bigl[(s\,\delta_i^\mu+t\sum_j\delta_j^\mu)\mxi^\mu
	+ \sum_{jkl}\Gamma_{ij}^{kl} \delta_j^\mu x_l^\mu {\bf
	J}_k^\mu \Bigr] \ ,
\end{equation}
where $s=1/a$ and
$t=-b/a(a+bK)$. Differential equations for the order
parameters can then be derived by the methods described in~\cite{ss}
and for the reduced dimensionality system we find,
\begin{eqnarray}
	\frac{\rd R}{\rd \alpha} & = & \eta\Bigl[ s\,\phi_s +
	t\,(\phi_s+(K-1)\phi_a) + v_R + w_R + z_R R\Bigr] \ ,
	\nonumber \\ \frac{\rd S}{\rd \alpha} & = & \eta\Bigl[
	s\,\phi_a + t\,(\phi_s+(K-1)\phi_a) + w_R + z_R S\Bigr] \ ,
	\nonumber \\ \frac{\rd Q}{\rd \alpha} & = & \eta\Bigl[
	s\,\psi_s + t\,(\psi_s+(K-1)\psi_a) + 2(v_Q + w_Q + z_Q
	Q)\Bigr] \nonumber \\ & & + \eta^2\Bigl[ s^2\upsilon_s +
	(2s\,t+t^2K)(\upsilon_s+(K-1)\upsilon_a)\Bigl] \ , \nonumber\\
	\frac{\rd C}{\rd \alpha} & = & \eta\Bigl[ s\,\psi_a +
	t\,(\psi_s+(K-1)\psi_a) + 2(w_Q + z_Q C)\Bigr] \nonumber \\ &
	& + \eta^2\Bigl[ s^2\upsilon_a +
	(2s\,t+t^2K)(\upsilon_s+(K-1)\upsilon_a)\Bigl] \ ,
\label{def_eom}
\end{eqnarray}
where,
\begin{eqnarray*}
	v_R & = & (\gamma_4+\gamma_6)(R-S)(\psi_s-\psi_a) , \\ w_R & =
	&
	(\gamma_4+\gamma_6)\Bigl[(R-S)\psi_a+S(\psi_s+(K-1)\psi_a)\Bigr]
	\ + \nonumber \\ & & \hspace{-2cm} (R+(K-1)S)\Bigl[
	(\gamma_2+\gamma_3+K\gamma_7)\psi_s +
	(2\gamma_8+\gamma_9+\gamma_{10}+K\gamma_{11})(\psi_s+(K-1)\psi_a)\Bigr]\
	, \\ z_R & = & (\gamma_1+\gamma_5K)\psi_s +
	(\gamma_2+\gamma_3+K\gamma_7)(\psi_s+(K-1)\psi_a) \ ,
\end{eqnarray*}
and $v_Q$, $w_Q$ and $z_Q$ are the same except that $R$ and $S$ are
replaced by $Q$ and $C$ respectively everywhere they appear
explicitly. Here, $\mbox{\boldmath $\gamma$}=[\gamma_i]$ is defined in
Eq.~(\ref{def_gam}) and we have defined,
\begin{eqnarray*} 
	& & \langle\delta_{i} y_{n} \rangle_{\{ \xi \}} =
\delta_{in}\phi_s + (1-\delta_{in})\phi_a \ , \quad \langle
\delta_{i}\delta_{j} \rangle _{\{ \xi \}} = \delta_{ij}\upsilon_s +
(1-\delta_{ij})\upsilon_a \ , \\ & & \langle\delta_{i} x_{j} +
\delta_{j} x_{i}\rangle_{\{ \xi \}} = \delta_{ij}\psi_s +
(1-\delta_{ij})\psi_a \ ,
\end{eqnarray*}
where $\delta_{ij}$ with two indices denotes the Kronecker delta and
brackets denote averages over the inputs. These averages can again be
calculated in closed form~\cite{ss}.

\subsection{Asymptotic dynamics}
\label{app_eom_asy}

The asymptotic dynamics for GD with an annealed learning rate have
recently been solved under the statistical mechanics formalism and the
optimal generalization error decay is known in this
case~\cite{lss}. Here we extend those results to NGD.

Following the notation in~\cite{lss} we define ${\bf u} =
(R-T,Q-T,S,C)^\tr$ to be the deviation from the asymptotic fixed
point. If the learning rate decays according to some power law then
the linearized equations of motion around this fixed point are given
by,
\begin{equation}
	\frac{\rd {\bf u}}{\rd\alpha} = \eta_\alpha\,{\bf M}\,{\bf u}
	+ \eta_\alpha^2\sigma^2{\bf b} \ ,
\label{eom_asy}
\end{equation}
where ${\eta_\alpha\,\bf M}$ is the Jacobian of the equations of
motion to first order in $\eta$ while the only non-vanishing second
order terms are proportional to the noise variance. For $T=1$ we find,
\begin{eqnarray*}
	{\bf M} & = & \frac{1}{\Delta} \left(\begin{array}{c c c c}
	c_1 & c_2 & -6(K-1)c_3 & 3(K-1)c_3 \\ -\frac{16}{\sqrt{3}}c_3
	& c_4 & -12(K-1)c_3 & 6(K-1)c_3 \\ 8\sqrt{3} & -4\sqrt{3} &
	c_5 & -9(K-1) \\ 16\sqrt{3} & -8\sqrt{3} & 2c_4 &
	-\frac{8}{\sqrt{3}}c_3 \end{array}\right) \ , \\ \Delta & = &
	\pi(25-16\sqrt{3}+K(8\sqrt{3}-9)) \ , \\ c_1 & = &
	2(32\sqrt{3}-41-K(16\sqrt{3}-9)) \ , \\ c_2 & = & (
	57-48\sqrt{3}+K(24\sqrt{3}-9))/2 \ ,\\ c_3 & = &
	2(2\sqrt{3}-6+3K) \ , \\ c_4 & = & 7-16\sqrt{3}+K(9+8\sqrt{3})
	\ , \\ c_5 & = & 16\sqrt{3}-43+K(27-8\sqrt{3}) \ ,
\end{eqnarray*}
and the two non-zero entries in ${\bf b}$ are,
\begin{eqnarray*}
	b_2 & = & \frac{
	\pi(38\sqrt{3}-66+K(51-30\sqrt{3})+K^2(6\sqrt{3}-9))}{(2+(K-1)\sqrt{3})(\sqrt{3}-2)^2}
	\ , \\ b_4 & = &
	\frac{-3\pi(7-4\sqrt{3}+K(2\sqrt{3}-3))}{(2+(K-1)\sqrt{3})(\sqrt{3}-2)^2}
	\ .
\end{eqnarray*}
The solution to Eq.~(\ref{eom_asy}) with
$\eta_\alpha=\eta_0/\alpha$ is,
\begin{equation}
	{\bf u}(\alpha) = \sigma^2{\bf V}{\bf X}{\bf V}^{-1}{\bf b} \
	,
\end{equation} 
where ${\bf V}^{-1}{\bf M}{\bf V}$ is a diagonal matrix whose entries
$\lambda_i$ are eigenvalues of ${\bf M}$. We have defined the diagonal
matrix {\bf X} to be,
\begin{equation}
	{\bf X}^{\mbox{\tiny diag}}_i =
	-\frac{\eta_0^2}{1+\lambda_i\eta_0}\left[\frac{1}{\alpha} -
	\alpha^{\lambda_i\eta_0}\alpha_0^{-(1+\lambda_i\eta_0)}\right]
	\ ,
\label{u_soln}
\end{equation}
where annealing begins when $\alpha=\alpha_0$. For natural gradient
learning we find two degenerate eigenvalues $\lambda_{1,2}=-1$,
$\lambda_{3,4}=-2$ and by substituting Eq.~(\ref{u_soln}) into a
first order expansion of the generalization error it is
straightforward to show $\eta_0=1$ to be optimal. In this case the
modes corresponding to $\lambda_{1,2}$ do not contribute to the
asymptotic generalization error and for all values of $T$ we find,
\begin{equation}
	\mbox{\large $\epsilon$}_{g} =
	-\frac{\sigma^2\eta_0^2K}{2\alpha(1+\eta_0\lambda_{3,4})} \ =
	\ \frac{\sigma^2K}{2\alpha} \ .
\end{equation}

\begin{references}

\bibitem{book} D.~Saad (ed), {\it Online learning in neural networks},
Publications of the Newton Institute (Cambridge University Press,
London, 1998)

\bibitem{amari} S.~Amari, Neural Comput. {\bf 10}, 251 (1998).

\bibitem{yang} H.~H.~Yang and S.~Amari, in {\em Advances in Neural
Information Processing Systems}, edited by M.~I.~Jordan, M.~J.~Kearns
and S.~A.~Solla (MIT Press, Cambridge, MA, 1998) Vol.~10, p~385.

\bibitem{yang2} H.~H.~Yang and S.~Amari, (unpublished, 1998).

\bibitem{rsa} M.~Rattray, D.~Saad and S.~Amari,
Phys. Rev. Lett. {\bf 81}, 5461 (1998).

\bibitem{cover} T.~Cover and J.~Thomas, {\em Elements of Information
Theory} (John Wiley \& Sons, New York, 1991).

\bibitem{orr} G.~B.~Orr and T.~K.~Leen in {\em Advances in Neural
Information Processing Systems}, edited by Mozer, Jordan and
Petsche (MIT Press, Cambridge, MA, 1997) Vol.~9, p~606. 

\bibitem{silvia} S.~Scarpetta, M.~Rattray and D.~Saad (unpublished, 1998).

\bibitem{ss} D.~Saad and S.~A.~Solla, Phys. Rev. Lett. {\bf 74},
4337 (1995); Phys.~Rev E {\bf 52}, 4225 (1995).

\bibitem{barber} D.~Barber, D.~Saad, and P.~Sollich,
Europhys. Lett. {\bf 34}, 151 (1996).

\bibitem{rb} P.~Riegler and M.~Biehl J.~Phys.~A {\bf 28}, L507 (1995).

\bibitem{biehl} M.~Biehl, P.~Riegler, and C.~W\"ohler,
J.~Phys.~A {\bf 29}, 4769 (1996). 

\bibitem{ws} A. H. L.~West and D.~Saad, Phys.~Rev.~E {\bf 56}, 3426 (1997).

\bibitem{sr} D.~Saad and M.~Rattray, Phys.~Rev.~Lett. {\bf 79}, 2578
(1997); M.~Rattray and D.~Saad, Phys.~Rev.~E {\bf 58}, 6379 (1998).


\bibitem{vicente} R.~Vicente and N.~Caticha, J.~Phys. A {\bf 30}, L599
(1997).

\bibitem{lss} T.~K.~Leen, B.~Schottky, and D.~Saad, in {\em Advances in
Neural Information Processing Systems}, edited by
M.~I.~Jordan, M.~J.~Kearns and S.~A.~Solla (MIT Press, Cambridge, MA,
1998) Vol.~10, p~301; Phys.~Rev.~E {\bf 59}, in press (1999).

\bibitem{am} S.~Amari and N.~Murata, Neural Comput. {\bf 5}, 140 (1993).

\bibitem{cybenko} G.~Cybenko, Math.~Control,~Signals and Systems {\bf 2},
303 (1989).

\bibitem{newton} M.~Rattray and D.~Saad, in~\cite{book}, p~185.

\end{references}
\end{document}